**Title:** Geometric Morphometrics approach for classifying children's nutritional status on out of sample data

**Authors:** Medialdea Laura*[1,2], Arribas-Gil Ana[2], Pérez-Romero Álvaro[2], Gómez Amador.[1]

1. Research, Development and Innovation Department, Action Against Hunger Spain.
2. Statistics Department. Universidad Carlos III de Madrid, Spain.

* email: lmedialdea@accioncontraelhambre.org

**Abstract:**
Current alignment-based methods for classification in geometric morphometrics do not generally address the classification of new individuals that were not part of the study sample. However, in the context of infant and child nutritional assessment from body shape images this is a relevant problem. In this setting, classification rules obtained on the shape space from a reference sample cannot be used on out-of-sample individuals in a straightforward way. Indeed, a series of sample dependent processing steps, such as alignment (Procrustes analysis, for instance) or allometric regression, need to be conducted before the classification rule can be applied. This work proposes ways of obtaining shape coordinates for a new individual and analyzes the effect of using different template configurations on the sample of study as target for registration of the out-of-sample raw coordinates. Understanding sample characteristics and collinearity among shape variables is crucial for optimal classification results when evaluating children's nutritional status using arm shape analysis from photos. The SAM Photo Diagnosis App© Program's goal is to develop an offline smartphone tool, enabling updates of the training sample across different nutritional screening campaigns.

1.  **Introduction:**

The classification of phenotypic differences among human groups is crucial for understanding adaptation to different environments and how factors like nutritional status influence these variations. Anthropological research has provided insights into the complex interplay between environmental conditions and cultural practices in shaping phenotypic diversity across human populations [1]. Moreover, studies in evolutionary biology have emphasized the importance of phenotypic variation as a reflection of adaptive responses to diverse ecological pressures [2]. Understanding the phenotypic manifestations of nutritional status is particularly relevant given its profound implications for human health and survival [3]. By elucidating the morphological adaptations that underlie variations in nutritional status among human populations, researchers can inform strategies for improving public health interventions and addressing global health disparities [4].

Geometric morphometric (GM) techniques, which use quantitative methods to study shape variation, provide detailed insights into these intraspecific variations. These methods can capture subtle morphological differences within a species by analyzing landmarks on the organism's structure, allowing for precise comparison of shape variations and their underlying causes [5,6]. GM offers a promising avenue for nutritional assessment, particularly in regions with limited resources. Traditional methods for evaluating infant and child nutritional status primarily rely on anthropometric measurements, which can be challenging to implement in low- and middle-income countries. Moreover, these methods often focus solely on linear measurements, overlooking critical information about body shape that is relevant to health. Recent studies have emphasized the limitations of traditional anthropometry in capturing complex aspects of body composition and shape [7]. GM provides an alternative approach to analyzing morphological variation by capturing both size and shape information using landmark-based techniques [6]. This allows for a more nuanced understanding of how nutritional status influences body morphology, which is particularly crucial during periods of rapid growth and development in early childhood [8]. Moreover, the advantage of digital health in managing malnutrition in low- and middle-income countries includes improved accessibility and efficiency. Digital health technologies enable remote diagnosis and monitoring, reducing the need for physical healthcare infrastructure, and facilitating timely interventions by healthcare providers. By incorporating geometric morphometric techniques into nutritional assessment protocols, researchers can gain valuable insights into the physiological impacts of malnutrition and better tailor intervention strategies to address the multifaceted dimensions of nutritional health [9]. In this line of work, the Severe Acute Malnutrition (SAM) Photo Diagnosis App Program©, led by Acción contra el Hambre, works to develop a smartphone application capable of applying GM techniques offline to identify the nutritional status of children aged 6 to 59 months from images taken of their left arm. Our goal is to support Health Systems in ensuring no malnourished child is overlooked by facilitating SAM nutritional screening with a reliable, user-friendly technological tool. This tool aims to: 1) establish efficient SAM screening systems for timely attention through individual follow-up and community nutrition surveillance; 2) digitalize and simplify epidemiological reporting on SAM, generating reliable data and indicators for informed decision-making; and 3) mobilize and engage the community in early SAM identification and referral for treatment. The program development relies on the collection of samples at varying times and in different contexts which can then be used to create a large reference sample from which a nutritional status classification rule can be derived. Such an approach has already been validated for severe acute malnutrition [9]. After validation on different populations, this rule would be used on the assessment of the nutritional status of new children through the smartphone application. The app would take a picture of a child's left arm and would automatically register the required landmarks and calculate the corresponding semilandmarks [10]. However, before proceeding to its classification, the obtention of the registered coordinates in the training reference sample shape space is required, and no standard techniques to perform this task are usually discussed in the literature.

When morphometry techniques based on linear measurements are used, it is common to construct discriminant functions that will indicate that one group is separated from the other for an estimated relationship between the different measurements used to construct the function. In this way, new individuals not belonging to the sample used to train the model can be evaluated by just calculating the same relationship between measurements. That is, the classification rule can be directly applied to new individuals.

When classifications are carried out using morphogeometric techniques, a classifier is generally built from the aligned coordinates of the sample studied, being the most commonly used linear discriminant analysis, although other approaches have also been tested, such as neural networks, logistic regression or support vector machine, among others [11–13]. Any chosen classification method should always be tested on data that has not been included in the model training stage [14]. In the GM field, the benchmark is to split data into training and test sets or leave-one-out individuals just after joint generalized Procrustes Analysis (GPA) of the whole dataset [15–20], although other alignment methods could also be used [21]. For example, in previous studies, this team designed a classifier for SAM applying a linear discriminant analysis to images of the left arm of a Senegalese sample of children aged between 6 and 59 months of age and tested it with leave-one-out cross-validation following this approach.

Although the combination of GM techniques with various methods for constructing classifiers has been extensively evaluated, and the theoretical procedures for assessing model performance are well systematized, the process for evaluating individuals not included in the training samples, also known as out-of-sample data, in real-world scenarios remains poorly understood. The problem lies in the fact that, in GM, classifiers are constructed not from the raw coordinates that define the landmark configurations but from transformations that utilize the entire sample's information. Typically, this involves Procrustes coordinates derived from GPA, but it could also be any set of aligned coordinates obtained with a different alignment method. However, it is not clear how this registration is applied to a new individual without conducting a new global alignment. This work proposes a methodology to evaluate out-of-sample cases from a classification model created from a training sample and evaluates its performance based on different choices of a template for registration of new individuals, as well as possible artifacts derived from the application of geometric morphometric methods.

2. Sample and methods:

2.1 Sample

A sample of images was taken from the left arm of 410 Senegalese girls (n= 206) and boys (n= 204) between 6 and 59 months of age. The data is made up of two datasets, one already used in previous studies [9] and another of 170 individuals collected between August and October 2023. Both datasets come from convenience sampling designed to find cases with SAM (n=202) and optimal nutritional condition (ONC, n=208) with equal proportions with respect to the nutritional status, age and sex of the individuals (Supplementary Table S1). The samples were collected in schools, nurseries and communities in Matam, one of the regions in the country with higher SAM prevalence. Selection criteria were as follows 1) children aged 6 to 59 months; 2) classified as either having ONC (MUAC between 135 and 155 mm or WHZ between -1 and +1 SD) or SAM (MUAC < 115mm or WHZ < -3SD); 3) absence of any physical malformation that could affect arm anatomy; 4) no medical complications other than the SAM condition and 5) a consent form signed by the legal guardians of every participant child. During the second sampling phase, an exclusion criteria was introduced: children with marks or scars in the arm that could eventually allow their identification. Caregivers of all children participating in the study gave written, informed consent and the study was approved by National Ethical Committee for Health Research (CNERS), Ministry of Health and Social Action, Dakar, Senegal (SEN 16/35, no. 164/MSAS/DPRS/CNERS and SEN 23/52, no. 232/MSAS/CNERS/SP, respectively), ensuring the study was conducted in accordance with the Declaration of Helsinki. Privacy data protection was ensured according to the Senegalese and European data protection laws.

To address the effect of age, two age groups were established according to the World Health Organization (WHO) guidelines for categorizing children under 5 years based on the method of measuring length or height recumbent (n=202 children aged up to 24 months) or standing (n=208 over 24 months) [22]. To ensure a representative distribution of all ages within these two groups, the study design included clustering infants and children between 6 and 59 months into five age ranges: 6–12 months, 13–24 months, 25–36 months, 37–48 months, and 49–59 months. Thus, the sample was equally distributed for age (up to and over 24 months) and sex in groups of around 100 children each (Supplementary Table S1). Regarding nutritional status, each group was subdivided in ONC or SAM condition, accounting for around 50 children in each sex/age group/nutritional status group.

**2.2 Methods**

**2.2.1. Anthropometric measurements and nutritional status determination**

To identify the nutritional status of the participants, anthropometric measurements of weight, height, and mid-upper arm circumference (MUAC) were registered. To ensure the accuracy of the measurements and the ultimate image capture if they accomplished the enrolling criteria, participants were undressed to their underwear or tight swimming costume bottom. Measurement equipment used for the first sampling included: a portable electronic scale Soehnle, calibrated to 0.1 kg for measuring weight; a portable infantometer (range 10–100 cm, precision: 0.1 cm) for measuring length in children under 87 cm or 24 months of age (closed intervals for both factors); and for children over 87 cm or 24 months of age, height was measured using a GPM anthropometer (range 10–230 cm, precision: 0.1 cm). MUAC was measured using a self-retracting, 0.7 cm-wide flat metal tape with a blank lead-in strip (precision: 1 mm). For the second sampling, similar certified equipment was used. Weight was measured with a portable electronic scale SECA 874, calibrated to 0.1 kg. A wooden portable enfant/child length/height measuring board, (range 0–129 cm to a 0.2 cm precision level) [23] was used to measure length and height. Finally, MUAC was measured through a self-retracting, 0.7 cm-wide, flat metal tape with a blank lead-in strip (with a precision level of 1 mm). Before starting each data collection session, calibration of scales and substitution of damaged measuring tapes was carried out. All measurements of weight (technical error 0.10 kg), length/height (technical error 0.15 mm) and MUAC (technical error 0.15 mm) were registered twice and only by one observer with an assistant for length/height measurement.

Just after registering the measurements, the R package "Anthro" provided by WHO [24] was used to calculate the weight-for-height/lenght z-score (WHZ) indicator commonly used to identify SAM together with MUAC, as well as other nutritional status indicators (descriptive analysis provided in Supplementary Table S2). Generated data were evaluated just after recording the measurements, and WHZ and MUAC nutritional indicators were estimated, considering SAM those children with WHZ < − 3 SD or MUAC < 125 mm (one or both criteria) and ONC between percentiles 15 and 85 for WHZ (between − 1 and 1 SD) and MUAC ≥ 135 (both criteria required). All caretakers of children participating in the study received a nutritional status assessment regarding their children by means of a written or oral report. Every child included in the study was given an identification code to ensure health information (registered in a database) and eventual image data privacy.

**2.2.2. Shape data acquisition.**

The methods for image registration, previously detailed for whole-body image capture [10], have been adapted in this study specifically for the capture of left arm images. Following the established methodology, four anatomical points were marked on the children's anterior body using dermatologically tested pens (Figure 1): shoulder (lateral acromion process), forearm (junction between the lateral epicondyle of humerus and the radial head), wrist (palmar midcarpal joint between the lunate and capitate bones), and armpit (intersection of the trunk contour line drawn from this point with the contour line drawn from the first maximum curvature identified in the arm). Children were positioned on a mat and anterior view images of their outstretched left arm were taken, held by two people to minimize movement. Up to 10 pictures in sequence were collected for each individual. The developed methodology accounted for the challenges of studying infants and children, who often tire quickly and may not cooperate due to their young age and limited understanding. Our approach, suitable for children from 6 months old, can potentially be adapted for younger ages based on research needs. The study involved adapting procedures to create a comfortable environment for children, families, and researchers, ensuring a calm and confident atmosphere. We engaged families and involved them throughout the data collection process. Participation was voluntary, and our intervention was crucial in providing valuable health knowledge to families in contexts with limited healthcare access.

A picture selection was performed from each individual's registered picture sequence and selected images were processed with Gimp 2.10 to remove shadow and orientation effects as well as to identify points marked on the left arm of the children and perform the geometric calculations required to replicate the previously established template [10]. Thus, geometric calculations were used to establish up to 16 semilandmarks on the body contour, based on four initial anatomical landmarks (Fig. 1). Lines connecting landmarks within the same region were drawn, with the forearm-to-wrist line serving as a reference for placing equidistant semilandmarks. These were divided by whole-number intervals informed by anatomical atlases and body proportion studies[9]. Parallel or perpendicular lines extended from these divisions to the body contour, and intersections were designated as semilandmarks. The method ensures homologous placement of semilandmarks across individuals, following validated methodologies for accurate body shape representation. Landmark and semilandmark configurations registration were then accomplished using tpsDig2 software[25].

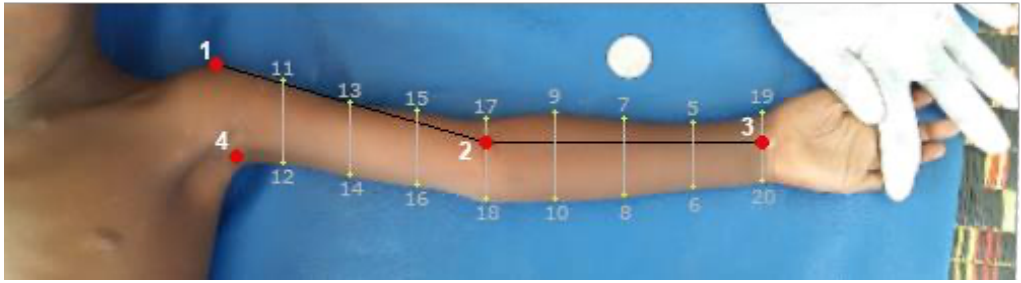

Figure 1. Landmark-based template of the left arm of a child in anterior view. Landmarks (red dots) identified and marked in the body of the child. Semi-landmarks (yellow dots) calculated to register child's arm shape from geometric calculations (grey lines) from landmark unions (black lines).

GM methods were used to visualize major shape changes as landmark-based deformations using thin-plate spline and to perform multivariate analyses on landmark coordinates. These analyses assessed the effects of relative size, sex, and age (continuous and grouped by under and over 24 months) on the children's left arm shape, stratified by nutritional status, using MorphoJ [26]. Procrustes ANOVA ($p < 0.05$) was performed before and after allometric correction to examine the effects of age (grouped), sex, and nutritional status (categorized as SAM and ONC) on both shape and size components (results to analysis provided in Supplementary Table S3). Additionally, a principal components analysis (PCA) of the shape variables before size correction, was performed to obtain a set of basis vectors representing the major trends in shape variation. The principal component (PC) scores for the first two principal components were displayed in a scatter plot, with 90% confidence ellipses used to group individuals according to age group and nutritional status. Shape differences were visualized through deformation grids, illustrating the variations in the objects' form.

### 2.2.3 Invariant shape analysis and alternative alignment methods
The current knowledge on the relationship between morphometric variations and nutritional status [9,10] relies on General Procrustes Analysis, as this is the standard preprocessing technique in GM.
However, given that the classification of new individuals under the alignment framework is challenging, and that this task can be much straightforward in the case of traditional morphometrics, we present here some alternative approaches to the study of the relationship between arm shape and nutritional status. We also consider some alternative alignment methods and compare preliminary in-sample classification results, to evaluate the differences between them.

In the study of shape, alignment-dependent methods, as discussed by [27], can introduce bias due to the arbitrary nature of the chosen coordinate system, whereas alignment free methods provide an invariant shape representation. Moreover, in the context of classification, invariant methods based on linear measurements have the advantage of building the classification rule as a function of the original observed coordinates, which can be straightforward applied to a new out-of-sample individual. That is, only the information available for the new individual is required in order to calculate the discrimination value, unlike for alignment-dependent methods, in which a training sample dependent transformation is required to be applied to the new individual before the classification rule can be computed.

The analysis of ratios of lengths or widths is a widespread approach in classical/multivariate morphometrics that allows to consider linear measurements for shape description and incorporate their discriminatory information into classification methods [28]. As a way of gaining understanding on the morphological differences between the arms shapes of children with SAM and ONC we have defined the following ratios: r1 = upper arm length / forearm length, r2 = upper arm width / forearm width, r3 = upper arm width / upper arm length, r4 = forearm width / forearm length. The corresponding lengths and widths are computed as the mean of the Euclidean distances between pairs of given landmarks: (1,17) and (4,18) for the upper arm length, (17,19) and (18,20) for the forearm length, (1,4), (11,12), (13,14), and (15,16) for the upper arm width, and (5,6), (7,8), (9,10) and (19,20) for the forearm width (see Figure 1 for reference).

The distribution of these variables across nutrional status, sex, age and subsamples is presented in Figure S1 of the Supplementary Material. Variables r3 and r4, that is the width to length ratios, seem to carry more information about the differences between the SAM and ONC than the ratios between lengths. The largest separation between groups are observed when considering the two different subsamples that compose the dataset.

In order to assess the discrimination power of the ratio variables with respect to nutritional status, the three classification methods used throughout the paper (linear discriminant analysis, logistic regression and k-nearest neighbors, see Section 2.2.4) have been applied on a leave-one-out classification strategy on the whole sample and in the two groups of age (below and above 24 months). The results are presented in Table S4 of the Supplementary Material.

Euclidean distance matrix analysis (EDMA [27]) is another invariant or alignment-independent shape method, consisting on calculating all pairwise Euclidean distances between landmarks for each individual. The individual intra-landmarks distance matrix is called form matrix (FM). From two form matrices, a form difference matrix (FDM) is defined as the ratio between the two, computed elementwise. Under the hypothesis that the two individual shapes are similar, all the ratios in the FDM should be similar, and so the ratio $T = max_{ij}(FDM_{ij})/min_{ij}(FDM_{ij})$ should be close to 1. If the FDM is calculated for each pair of individuals, the T value can be used to define and inter-individual distance matrix (by considering the values $\log T$). From it, multidimensional scaling (MDS) techniques are used for visualization and classification purposes.

The FDM can also be calculated between two estimated mean FMs coming from two groups of landmark configurations. In that case, the distribution of T under the hypothesis of no shape differences between the groups can be approximated via bootstrap to conduct a global shape difference test between the two groups.

A local test can also be performed if instead of looking at the T value, bootstrap is used to build confidence intervals for each of the FDM elements. Confidence intervals not containing the value 1 indicate significant differences in the corresponding FDM element, that is, in the distance between the corresponding pair of landmarks.

The results of the EDMA analysis (included in Section B2 of the Supplementary Material) indicate that there exist both global and local differences between the mean form matrices of the SAM and ONC groups. Specifically, the local differences arise mainly in the distances between landmarks accounting for the arm width, and they are larger for the upper arm width than for the forearm width. This is consistent with current knowledge about SAM, since one of the standard screening methods [22] is to measure the mid-upper arm circumference (MUAC), which is the arm circumference at approximately the line between semilandmarks 13 and 14 (see Figure 1). Regarding the ability to separate individuals in terms of their nutritional status, classification based on EDMA provides similar results to those obtained with ratios of lengths and widths (Table S5 in the Supplementary Materials).

In order to compare both EDMA and ratio analysis with alignment-based methods for the classification of nutritional status, we have chosen three well known alignment methods: generalized Procrustes analysis (GPA), weighted GPA (WGPA) [21], and square root velocity framework (SRVF)[29]. As illustrated by Taheri and Schulz [30], different alignment methods can yield significantly different outcomes, as each method is designed to minimize a specific distance between landmark configurations (see Figure S4, Supplementary Material) . In the classification context, we may choose an alignment method in terms of the classification results obtained from the corresponding aligned coordinates. Nevertheless, for out-of-sample classification, any alignment method will encounter the challenge of addressing non-aligned coordinates of new individuals. In essence, the classification rule is derived from a reference sample of aligned coordinates, requiring that any new individual, for whom only raw coordinates are available, be classified using this rule. Consequently, a pairwise alignment procedure to align the new individual with some reference target must be performed as a preliminary step. This is at the core of the methodology proposed in this paper. Yet, before introducing the new proposal for the alignment and classification of new individuals, we want to understand if the choice of an alignment method can significantly impact the nutritional status classification performance for our data set. At this exploratory stage, their in-sample classification performance will be used as a proxy of out-of-sample classification behavior.

The results are presented in the Supplementary Materials (Table S5), with the best classification results achieved using SRVF, which appears to be robust against allometric regression (size correction) and potential landmark collinearity.

However, the best scenario performance across all three methods is quite similar (accuracy of around 0.92-0.93). In all cases, it is significantly better than the best classification results obtained with either ratio analysis (accuracy of 0.70) or EDMA (accuracy of 0.75). Therefore, we consider that the classification power obtained in this case for alignment-based methods is superior and justifies the development of an out-of-sample approach for the application of the classification rule to new individuals.

Also, given that morphometric variations due to nutritional status were originally studied for a subset of this dataset relying on GPA [9], we opted to use GPA as the alignment method to demonstrate our methodology throughout the paper. Nevertheless, our approach is adaptable to any alignment method, as explained and illustrated through simulation in Section 2.2.6.

### 2.2.4. Classification methods

Linear discriminant analysis (LDA) is the gold standard classification method in geometric morphometrics [26,31,32]. It is a simple method that does not require the tuning of hyperparameters. This was the method used to predict the nutritional condition of a subgroup of children from the sample of this study in [9]. In this work, we compare its performance in relation with other standard methods in classification such as logistic regression (LR) and *k*-nearest neighbors (kNN). LR is expected to provide very similar results to LDA [14]. For kNN, which classifies an individual in the majority class of its closest *k* observations, the hyperparameter *k* needs to be chosen in advance. In this work, and just for the sake of comparability with the LDA benchmark, we adopt an oracle perspective by selecting the value of *k* as the one that maximizes the global accuracy in the in-sample classification of the whole sample (training sample equal to test sample), before applying it to out-of-sample leave-one-out classification (see Section 3.2).

### 2.2.5. Dealing with collinearity

The 20 landmark and semilandmark configuration selected for this study consists of semilandmarks calculated from one or two landmarks. The semilandmarks recorded on the wrist and forearm (two in each case) were estimated from a single landmark (wrist or forearm, respectively) as a perpendicular from that point to the orientation line used for semilandmarks calculations (see Figure 1). Therefore, when considering the k×2 Procrustes coordinates (or corresponding allometric residuals) as predictive variables in a classification model, 4 of them are linearly dependent on the remaining ones. Specifically, the semilandmarks that define the contour of the wrist and forearm (k=4) are calculated directly from the two landmarks recorded at these anatomical points. To avoid collinearity, which might result in numerical instability and poor classification power, especially when using linear discriminant analysis, LDA, or logistic regression, LR [14] we propose to remove these two landmarks, therefore resulting in a model of 18 landmarks/semilandmarks and 36 classification variables. A common approach to dealing with collinearity in LDA and other classification methods is to perform a dimension reduction technique previous to classification, such as principal component analysis [33]. However, in terms of interpretability, since the source of linear dependence is known, we prefer in this case to just remove some of the original landmarks.

### 2.2.6. Out-of sample classification for alignment-dependent methods

In the context of an alignment-dependent shape analysis method, the classification methods described in the previous section are built on the aligned coordinates (or the corresponding residuals from the allometric regression of shape on size described in [10]) of the training sample. Although we could use any alignment method, for the sake of clarity in the following description we illustrate the methodology using GPA as alignment method.
If the configuration matrices of the *n* individuals of the sample are $X_1, \ldots, X_n$, we will denote by $X_1^P, \ldots, X_n^P$ the Procrustes coordinates obtained after applying a full generalized Procrustes analysis (fGPA) (see [21]). We will also denote by $X_1^{P\ res}, \ldots, X_n^{P\ res}$ the residuals from the allometric regression of the Procrustes coordinates on the logarithm of the centroid size. Note that for each individual *i*, $X_i$, $X_i^P$ and $X_i^{P\,res}$ are *k×2* matrices, where *k* is the number of landmarks. Then, the classification rule defined on the training sample is built either on $X_1^P, \ldots, X_n^P$ or $X_1^{P\,res}, \ldots, X_n^{P\,res}$.

When the aim is to classify a new out-of-sample individual, its Procrustes coordinates are not available, since it did not belong to the original sample. Therefore, the first step before classification is to obtain the Procrustes coordinates of the new individual (and possibly the corresponding residuals from the allometric regression). One option would be to conduct fGPA on the new sample defined by $X_1, \ldots, X_n, X_{new}$, where $X_{new}$ represents the configuration matrix of the new individual. However, this will lead to Procrustes coordinates $X_1^{P*}, \ldots, X_n^{P*}$, $X_{new}^{P*}$ for which the first *n* might not be equal to the original Procrustes coordinates $X_1^P, \ldots, X_n^P$. Therefore, the classifying rule based on the original Procrustes coordinates might no longer be valid on $X_1^{P*}, \ldots, X_n^{P*}, X_{new}^{P*}$. Consequently, we need to consider the obtention of the Procrustes coordinates for the new individual independently of those of the original sample, which must remain fixed.

Another possibility is to perform a full ordinary Procrustes analysis (fOPA) of the new individual raw coordinates to some target configuration obtained from the Procrustes coordinates of the reference sample. This target could be the mean Procrustes shape, but it could also be a different central pattern. In fact, when the chosen target is the mean shape, $\bar{X}^P = \frac{1}{n}\sum_{i=1}^{n} X_i^P$, the minimization problem to solve the fOPA of $X_{new}$ to $\bar{X}^P$ is equivalent to finding the best transformation of $X_{new}$ (rotation, translation, scale) that globally minimizes the sum of square Euclidean distances to all Procrustes coordinates in the reference sample. However, in the presence of atypical individuals in the original set, or just a very heterogeneous sample, more robust choices can be considered, such as defining the target as a "median shape". Analogously to what is done to obtain $\bar{X}^P$, a pointwise median configuration is defined as the configuration whose coordinates are given by the median of each of the Procrustes coordinates. However, this would not result in one of the configurations in the individual sample. If the aim is to obtain a central pattern which is one of the observations of the sample, in the spirit of the median of a univariate sample, a different notion of median has to be considered.

We will refer to this alternative median as *functional median* which is based on the definition of the median curve in the field of functional data, where a curve is considered to be the (sample) median if it maximizes the *depth* within a given set of curves defined on the same domain [34]. A statistical depth measure is a term for a measure that quantifies how central an observation is within a sample, establishing a central-outwards ordering that generalizes the natural order in the real line. In functional data analysis, where observations are curves or surfaces, several definitions of functional depths exist, but we will use the *modified band depth* [34], although alternative depth measures may yield similar results. From the order established by the depth values, the median curve is the deepest curve, intuitively the curve most surrounded in a sample of curves, i.e., it is the most central curve. Notice that 2-dimensional Procrustes configurations can be considered as functional data objects. Indeed, the sequence of (x,y)-coordinates at the different landmarks and semilandmarks can be thought of a discretely observed version of the continuous arm contour in $R^2$. Therefore, we will obtain the modified band depth of the Procrustes configurations and use the deepest one (just functional median in the following) as target for fOPA.

In Figure 2 we illustrate the differences between the mean, the pointwise median and the functional median in two different sets of Procrustes coordinates.

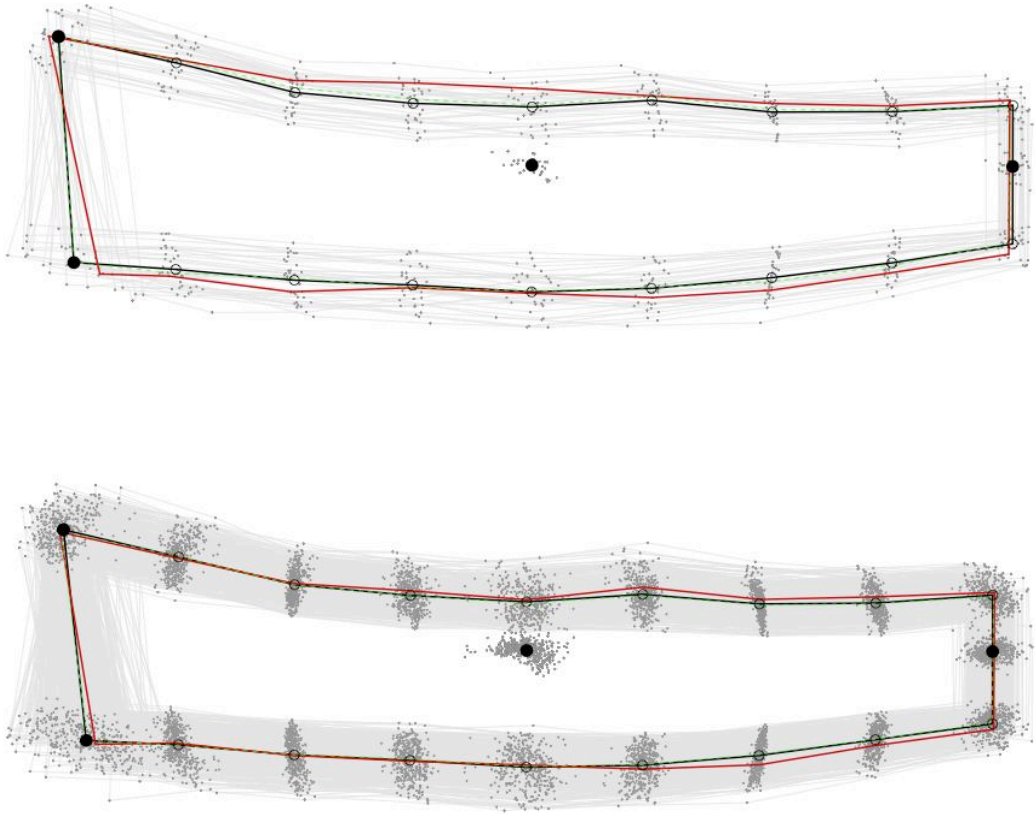

Figure 2: Procrustes coordinates (dark gray dots joined by light gray lines), mean (solid black), pointwise median (dashed green) and functional median (solid red). Top: 36 selected individuals. Bottom: whole sample (410 individuals). Landmarks are represented by solid bullets and semilandmarks by empty bullets and they are all located in the graphic at the mean position of the corresponding coordinates. In the two graphics, the mean and pointwise median are almost indistinguishable. However, the relation between the functional median and the mean is different depending on the sample. In the small sample there are visible differences between the two central patterns whereas in the whole sample these differences are almost negligible.

Finally, considering these two possible targets for registration of new individuals, the mean and functional median of the Procrustes variables, we now describe the step-by-step procedure that will allow us to classify and assess the classification performance of different methods on out-of-sample individuals.

*Algorithm 1: out-of-sample classification of a new individual*

1. Consider the *n* configuration matrices of the reference sample $X_1, \ldots, X_n$ and conduct fGPA to obtain $X_1^P, \ldots, X_n^P$.
   1a. For allometric correction, perform allometric regression on $X_1^P, \ldots, X_n^P$ to obtain the corresponding residuals, $X_1^{P\,res}, \ldots, X_n^{P\,res}$.
2. Train the LDA/LR/kNN classifier on the sample obtained in 1.
3. Calculate the reference target, RT, either from $X_1^P, \ldots, X_n^P$ or $X_1^{P\,res}, \ldots, X_n^{P\,res}$. We consider two possibilities for RT, either the mean or the functional median of the coordinates.

For a new individual with configuration matrix given by $X_{new}$:

4. Conduct fOPA of $X_{new}$ into RT to obtain $X_{new}^P$.
   4a. For allometric correction, obtain $X_{new}^{P\,res}$ from $X_{new}^P$ by means of the allometric regression coefficients estimated in 1a.
5. Apply the classification rule obtained in 2 to $X_{new}^P$ or $X_{new}^{P\,res}$.

Notice that the proposed algorithm differs from incorporating the new individual into the reference sample before obtaining a global alignment. As previously explained, in the context of children malnutrition detection with the offline mobile application SAM Photo Diagnosis App©, a fixed study reference sample is used to build a classifier which remains unchanged on every evaluation of the app on a new child.

If the aim is to evaluate the performance of algorithm 1 on a real or simulated data set, a leave-one-out cross validation strategy should be used.

*Algorithm 2: Leave-one-out cross validation for out-of-sample classification of new individuals*
For i=1, ..., n:
1. Consider the n-1 configuration matrices $X_1, \ldots, X_{i-1}, X_{i+1}, \ldots, X_n$ and conduct fGPA to obtain $X_1^P, \ldots, X_{i-1}^P, X_{i+1}^P, \ldots X_n^P$.
   1a. For allometric correction, perform allometric regression on $X_1^P, \ldots, X_{i-1}^P, X_{i+1}^P, \ldots X_n^P$ to obtain the corresponding residuals, $X_1^{P\,res}, \ldots, X_{i-1}^{P\,res}, X_{i+1}^{P\,res}, \ldots X_n^{P\,res}$.
2. Train the LDA/LR/kNN classifier on the sample obtained in 1.
3. Calculate the reference target, RT, either from $X_1^P, \ldots, X_{i-1}^P, X_{i+1}^P, \ldots X_n^P$ or $X_1^{P\,res}, \ldots, X_{i-1}^{P\,res}, X_{i+1}^{P\,res}, \ldots X_n^{P\,res}$. We consider two possibilities for RT, either the mean or the functional median of the coordinates.
4. Conduct fOPA of $X_i$ into RT to obtain $X_i^P$.
   4a. For allometric correction, obtain $X_i^{P\,res}$ from $X_i^P$ by means of the allometric regression coefficients estimated in 1a.
5. Apply the classification rule obtained in 2 to $X_i^P$ or $X_i^{P\,res}$.

Notice that the difference between the previous algorithm and an in-sample classification strategy is that for the later, all *n* aligned configurations $X_1^P, \ldots, X_i^P, \ldots X_n^P$ are obtained just once and leave-one-out cross validation is conducted by removing each of them at a time and training the classifier on the remaining *n-1*.
If a different alignment method other than generalized Procrustes analysis is used, the previous algorithm is still valid by replacing fGPA by the chosen global alignment method and fOPA by its pairwise version. In this case, $X_i^P$ will simply refer to the aligned coordinates of the i-th individual obtained with the chosen method. In this situation, it might be desirable to consider a different pattern for alignment of a new individual. For instance, for SRFV alignment, one could consider the Karcher mean or median [29], but different targets could be defined to replace the sample mean and functional median in the previous algorithms.

In order to demonstrate the behavior of the proposed methodology, a simulation study has been designed to illustrate the use of the mean and the functional median as reference target, different classifiers and different alignment methods. The specifications and results are presented in Section D of the Supplementary Materials. One interesting conclusion is that if the groups exhibit differences in mean shape, the use of the sample mean as reference target yields better results. On the other hand, if the two groups have the same mean shape but are different in variation, the functional median will provide better classification results, at least for alignment with GPA.

Statistical analyses were performed in R statistical software (version 4.3.2). The 'anthro' package (version 1.0.1) was used for nutritional indicators calculation, the 'shapes' package (version 1.2.7) was used for generalized and ordinary Procrustes analysis, and for weighted Procrustes analysis, the 'fdasrvf' package was used for square root velocity framework alignment (version 2.0.3), the 'geomorph' package was used for allometric regression (version 4.0.5), the 'roahd' package (version 1.4.3) was used for functional median calculation, the 'EDMAinR' package was used for EDMA (version 0.3-0), the 'MASS' package (version 7.3-60) was used for linear discriminant analysis, and the 'stats' package, included in the base R distribution, was used for logistic regression.

## 3. Results:

### 3.1. Morphogeometric variation of the left arm

A PCA was conducted to explore major arm shape changes and to examine their relationship with the factors of interest of the study: age groups and nutritional status. Sex was excluded from this analysis, as it did not have a significant effect on arm shape variation (Suppl. Table S3). A total of 36 PCs were extracted from the analysis before carrying out the size correction, where PC1 explained 50.013% of the total variance, and PC2 31.458%, both accounting 81.471%. The positive forms along PC1 are generally thicker than the mean shape, showing a centrifugal expansion at all landmarks marking the arm contour. Additionally, a shortening of the upper arm region is observed (Fig. 3a). This shape corresponds to that of children in optimal nutritional condition up to 24 months of age. In contrast, the negative forms along PC1 appear narrower than the mean shape, with reduced dimensions at the landmarks outlining the arm contour. An elongation of the upper arm region is also noted. This morphology is associated with children over 24 months suffering from SAM.

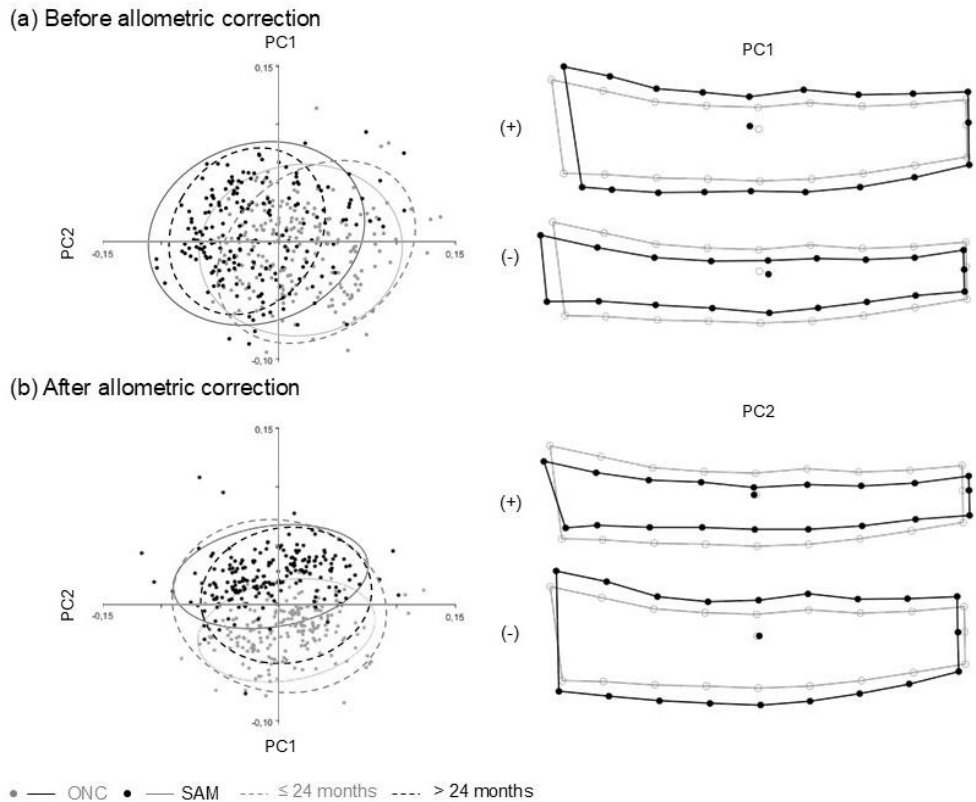

Figure 3. Shape variation in the left arm of infants and children aged 6–59 months. Principal component analysis was conducted both with (a) and after correcting for (b) the allometric effect, highlighting the main shape changes along PC1 and PC2, respectively (mean shape shown in grey). The PC1 vs. PC2 plots display the sample distribution based on nutritional status: optimal nutritional condition (ONC, represented by grey ellipses and dots) and severe acute malnutrition (SAM, represented by black ellipses and dots). Age groups are also distinguished, with children up to 24 months shown by a grey dashed ellipse and those over 24 months by a black dashed ellipse.

After performing the allometric correction, again 36 PCs were extracted from the PCA, where PC1 explained 45.133% of the total variance, and PC2 30.674%, both accounting 75.807%. In this case, shape differences associated with our study factors are observed in PC2 (Fig. 3b). Positive forms display thin arm morphologies, with a marked centripetal retraction of the semilandmarks that outline the arm contour, aligning with the location of the main muscle and adipose tissue groups in the arm. This morphology is associated with SAM condition, predominantly represented along the positive region of PC2. Negative forms along PC2 correspond to arms with greater width across the entire contour structure, similarly capturing the form of major muscle and adipose tissue groups. This shape aligns with that of children with ONC, predominantly represented in the negative region of PC2. Age groups now appear overlapped; indeed, the morphology of the up-to-24-months group fully encompasses that of the over-24-months group. Therefore, the effect of age is mitigated after applying the allometric correction.

## 3.2. Classification results: whole sample

We now present the classification results obtained using out-of-sample leave-one-out cross validation on the whole sample as described in Algorithm 2, Section 2.2.6. The classification performance of the four different strategies (with or without allometric correction/mean or functional median for fOPA) combined with the three classification methods is summarized in Table 1. The same procedure was carried out after removing the coordinates corresponding to landmarks 2 and 3 to avoid collinearity (Table 1). We can observe the effects of size correction and the elimination of linearly dependent coordinates on classification. Size correction affects the three classification methods, with the best results being obtained when removing the allometric effect. Collinearity also influences the performance of both LDA and LR, but numerical instability specifically impacts the performance of these methods when based on fOPA to the functional median, which should not be used in that context. When removing collinearity and the allometric effect, both fOPA to the mean or the functional median are very similar with each of the three classification methods. In particular, the results of LDA or kNN are slightly superior to those obtained with LR.

Table 1. Leave-one-out classification results on the whole sample (410 individuals) by Linear discriminant analysis (LDA), Logistic regression (LR) and *k*-nearest neighbours (kNN). Acc stands for Accuracy, Sens for Sensitivity for SAM and Spec for Specificity for SAM. Numbers in bold represent accuracy values larger than 0.90.

| Class. Method | Allometric effect | 20 landmarks - 40 variables | | | | | | 18 landmarks - 36 variables | | | | | |
| --- | --- | --- | --- | --- | --- | --- | --- | --- | --- | --- | --- | --- | --- |
| | | fOPA to mean | | | fOPA to functional median | | | fOPA to mean | | | fOPA to functional median | | |
| | | Acc | Sens | Spec | Acc | Sens | Spec | Acc | Sens | Spec | Acc | Sens | Spec |
| LDA | Before size correction | 0.7390 | 0.8564 | 0.6250 | 0.4293 | 0.4554 | 0.4038 | 0.7585 | 0.8762 | 0.6442 | 0.7951 | 0.8020 | 0.7885 |
| | After size correction | **0.9146** | 0.9455 | 0.8846 | 0.4512 | 0.4653 | 0.4375 | **0.9268** | 0.9604 | 0.8942 | **0.9244** | 0.9208 | 0.9279 |
| LR | Before size correction | 0.7317 | 0.8564 | 0.6106 | 0.4220 | 0.4406 | 0.4038 | 0.7415 | 0.8762 | 0.6106 | 0.8000 | 0.8168 | 0.7837 |
| | After size correction | 0.8805 | 0.9109 | 0.8510 | 0.4366 | 0.4505 | 0.4231 | 0.8951 | 0.9307 | 0.8606 | **0.9098** | 0.9109 | 0.9087 |
| kNN | Before size correction | 0.7146 | 0.6139 | 0.8125 | 0.7146 | 0.6089 | 0.8173 | 0.7146 | 0.6238 | 0.8029 | 0.7073 | 0.6238 | 0.7885 |
| | After size correction | **0.9122** | 0.9257 | 0.8990 | 0.9146 | 0.9257 | 0.9038 | **0.9244** | 0.9356 | 0.9135 | **0.9220** | 0.9307 | 0.9135 |

## 3.3. Classification results by age and sex

Given the performance results obtained in the previous section, we now present the classification results by age and sex using only one of the best performing methods, namely, LDA on the 18-landmarks configuration and with the mean as reference target for the fOPA of out-of-sample individuals. Although best results are consistently obtained after size correction, we also present the results before size correction for reference. The classification results are summarized in Table 2. It can be observed that the classification results in the group of 0-2 years old are worse than those obtained in the whole sample. This is in line with what had been observed in [9]. In the group of the older children (2 to 5 years old), the obtained accuracy is only slightly higher than in the whole sample. In terms of sex, the classification results of both girls and boys separately are similar to the ones obtained in the whole sample.

| Allometric effect | By age group | | | | | | By sex | | | | | |
|---|---|---|---|---|---|---|---|---|---|---|---|---|
| | < 24 months (n=202) | | | ≥ 24 months (n=208) | | | Female (n=206) | | | Male (n=204) | | |
| | Acc | Sens | Spec | Acc | Sens | Spec | Acc | Sens | Spec | Acc | Sens | Spec |
| Before size correction | 0.7822 | 0.8100 | 0.7549 | 0.8269 | 0.9020 | 0.7547 | 0.7961 | 0.9010 | 0.6952 | 0.7549 | 0.8713 | 0.6408 |
| After size correction | **0.8663** | 0.8400 | 0.8922 | **0.9279** | 0.9510 | 0.9057 | **0.9233** | 0.9604 | 0.8857 | **0.9118** | 0.9604 | 0.8641 |
| Allometric effect | By age group and sex | | | | | | | | | | | |
| | < 24 months Females (n=102) | | | ≥ 24 months Females (n=104) | | | < 24 months Males (n=100) | | | ≥ 24 months Males (n=104) | | |
| Before size correction | 0.7941 | 0.8000 | 0.7885 | 0.7885 | 0.8627 | 0.7170 | 0.7000 | 0.8000 | 0.6000 | 0.8077 | 0.9020 | 0.7170 |
| After size correction | **0.8529** | 0.7800 | 0.9231 | **0.8846** | 0.9608 | 0.8113 | **0.8400** | 0.8800 | 0.8000 | **0.8654** | 0.9216 | 0.8113 |

Table 2. LDA leave-one-out classification results by age and sex and age/sex, when the target for fOPA of out-of-sample individuals is the Procrustes mean. Acc stands for Accuracy, Sens for Sensitivity for SAM and Spec for Specificity for SAM. Numbers in bold represent highest accuracy values for each subsample.

**Discussion:**

MG techniques have generated evidence related to many research questions concerning the morphometric variation of biological structures, such as their quantification and visualization in relation to different taxa [35–37], geographic distribution [16,38–40], various factors [9,10,41,42] and their relationship with phylogeny [43,44], among others. However, despite considerable advances in the complex capture and analysis of shape using this set of techniques, applying the acquired knowledge to new individuals that were not part of the study sample remains challenging.

One of the main reasons that prevent researchers from approaching out-of-sample solutions, is that morphometric traits can vary significantly across different populations due to genetic, epigenetic, environmental, and cultural factors as well as secular trends [16,45–47]. Thus, some authors suggest that the problem may lie in the fact that classifiers trained on one population may not accurately reflect the morphological variations present in another population, leading to reduced classification accuracy when applied to out-of-sample data [16,38,48]. Nevertheless, recent globalization, climate change and migratory trends has led to the movement of hundreds of thousands of people worldwide, resulting in extensive population mixing. This is creating a new debate about the suitability of adapting these tools to the current situation, and it is questioning whether some of the knowledge generated from past populations, which presumably experienced greater geographic isolation that preserved their phenotypic characteristics, is still applicable to contemporary populations. Numerous researchers have actively pursued and suggested the development of robust techniques designed to minimize dependence on the specific population sample used for their construction, such as the use of apportionment of variance for the combination of samples from different populations [38], the automatization of landmark acquisition to minimize the observer error across samples [15] or the incorporation of a high probability threshold allowing for "indeterminate results" in classification [15,46]. However, at this point, the discussion focuses more on how to construct and analyze study samples to achieve satisfactory classification results for those samples, rather than on how to apply the results to individuals who were not part of the original samples. Cross-validation is the gold standard when discussing generalization of classification results to new individuals or samples. Nonetheless, in the context of GM in which shape is analysed after alignment of the landmark configurations, to the best of our knowledge cross-validation has always been studied through simulation on variables that are already in the shape space, that is, for which no Procrustes analysis is required [49,50], or on real data sets for which the whole sample is split into training and test sets only after all sample dependent preprocessing steps (GPA and possibly dimensionality reduction) have being jointly conducted [17,19]. In this work, we address the classification of new individuals with a rule previously obtained on a full Procrustes shape space computed from a training sample to which they did not belong. This is motivated by the aim of using the offline mobile application SAM Photo Diagnosis App© for the detection of potentially critical nutritional status on children. We propose two ways of projecting the out-of-sample individual into the training shape space, including considerations about sample heterogeneity, considering that the data in our study comes from two different data sources from the same population. In the future, with the incorporation of the aforementioned precautions to deal with samples of different population origins, this will allow to contribute to the current debate on the composition of reference samples and their extended use and application in contemporary human populations.

Sample size is another constraint when analyzing morphogeometric data [49,51–53]. Whereas some estimation procedures might be robust to small samples sizes, group comparisons have been shown to be extremely affected by the lack of a large enough number of individuals with respect to the number of variables. Moreover, in the classification problem, the fact of using imbalanced data, that is, a sample with an uneven distribution of cases across the different classification groups, may affect the overall method's performance even when the global sample size is large [50]. It is worth noticing that in our study the sample size is large enough (410 individuals globally, and around 100 when considering sex-age groups) and large in comparison with the total number of variables (40 coordinates from 20 2-dimensional landmarks). Also, the nutritional groups are balanced across the sample, with 208 children in the optimal nutritional class and 202 in the severe acute malnutrition class.

Moreover, high-dimensional morphometric data commonly used to register the shape of biological structures, with numerous variables representing shape coordinates, can lead to overfitting where the classifier performs well on training data but poorly on new data. Indeed, not only the sample size is a constraint here, as previously mentioned, but also the dependency among shape variables leading to multicollinearity, even when the sample size is large with respect to the number of coordinates. Common approaches to cope with it include extracting a reduced number of principal components from the landmark variables or the use of regularization techniques [53,54]. In our study, collinearity was clearly identified from the landmarks design, and could be simply dealt with by removing two redundant landmarks. Although LDA can perform well in some settings even when dealing with linear dependent variables (through the use of a pseudo-inverse for the covariance matrix), the results that we obtain by removing redundant variables are best in all classification scenarios.

Differences in data acquisition methods, such as imaging techniques or landmark placement, can introduce inconsistencies. These discrepancies may not significantly impact within-sample classification but can lead to substantial errors when applying the classifier to out-of-sample data collected under different conditions [55]. Morphological traits often interact in complex ways that are not fully captured by the models. Simplified models may fail to account for these interactions, reducing their robustness when applied to new individuals with different trait combinations [56]. Overall, the challenge lies in ensuring that the classifier captures the underlying biological and morphological principles that are consistent across different populations and not just the specific characteristics of the training sample. Moreover, it is well known that the number and type of structures studied have a more significant impact on the overall results of an analysis than the type of data – classical or GM-based metrics – derived from them [38]. Hence, it is essential to properly design the configuration of landmarks to accurately capture the shape of the biological structure under analysis, in this case, the left arm. Since we aim to establish an interpretable classification rule based on these variables, we have decided to retain this configuration of landmarks as much as possible, eliminating bias from collinearity, rather than performing a variable reduction (e.g., PCA) that is difficult to interpret. This approach has yielded excellent classification results, allowing us to establish a method for selecting landmarks that maintains interpretability and ensures usability in classification.

Regarding the two methods proposed to obtain the shape coordinates of new individuals, our results show that once the undesirable effects of collinearity are avoided, the performance of using the mean or functional median as reference target for fOPA is very similar in the sample of study. This is not surprising at the view of Figure 2 (bottom), where both central patterns are compared in the whole sample and are shown to be very similar. However, in smaller samples with higher heterogeneity or possible outlying profiles, it might be necessary to assess their influence in the mean profile and consider possible alternatives for fOPA of new individuals, such as the functional median. This has been illustrated with a simulation study in which the classification performance when using the functional median is better than that of the sample mean if the groups are similar in mean shape but different in variation.

Regarding the comparison between the different classification methods, kNN has been shown to obtain similar results to LDA. However, it is worth noticing that kNN requires the use of a hyperparameter, $k$, the number of nearest neighbors. In this work, this number has been selected previously in terms of the method's classification performance on the whole sample (before applying it to leave-one-out classification). In practice, this number could also be estimated from a validation sample and be kept fixed for every app use on the classification of a new individual. Nonetheless, from a practical point of view, LDA has a second advantage with respect to kNN. Notice that the smartphone App to be designed needs to be used off-line on low performance devices. In this sense, the application of the LDA classification on a new individual only requires the access to the reference target and the classification rule (which is a linear rule in terms of the group means and common estimated covariance matrix). That is, the app only needs to store four low dimensional objects and conduct a simple calculation, besides of the fOPA of the new individual. On the other hand, the implementation of kNN requires the storage of all individuals' coordinates and class labels, and the calculation of $n$ Euclidean distances, in addition to the fOPA of the new individual. Not only this involves higher storage and calculation requirements, but it can be problematic in terms of data privacy since the whole reference sample data should be available locally on each device in which the app is installed.

As limitations for this study, given that the classifiers trained might be biased towards the population of Senegalese children aged 5 to 59 months, it would be prudent to evaluate their performance on a different population of children, possibly from another country perhaps exhibiting a presumably different morphology. Furthermore, the mean and median registration methods perform similarly in our study because all training shapes were standardized using Procrustes coordinates and represent the same type of shape (an arm captured in the same position/from the same perspective), despite the fact of dealing with heterogeneity across the two different subsamples, as captured by invariant shape analysis methods. However, if the sample included different shape types and was more heterogeneous (from different approaches to picture taking, for instance), the median shape registration method might yield better results. Finally, although the sample size in our study is adequate for discussing observed morphological patterns related to ontogeny, sexual dimorphism, nutritional status variation, and population variability within the field of GM, it limits the use of more complex classification models in the computational domain. These models have numerous parameters and high capacity, necessitating extensive data to ensure robust performance and generalizability. However, simple classification rules are less prone to overfitting, which is a significant risk with small datasets. They require fewer computational resources and are faster to train, making them more practical for real-time applications or resource-constrained environments. Additionally, simple models are easier to interpret and validate, providing clearer insights into the data and facilitating better understanding and communication of the results. This interpretability is crucial for applications requiring transparency and explainability, such as clinical or diagnostic tools [57]. Moreover, the goal is to train computationally efficient models for the SAM Photo diagnosis app, designed to work offline and use minimal memory. Therefore, simple classification rules were selected.

**Conclusions:**

This work proposes a methodology to classify test data not included in the training sample used to infer/learn a classification model under alignment methods, considering the characteristics of data generated from GM techniques. Two methods are suggested for positioning the test data in the morphospace generated by the registered training data: pairwise alignment to the mean or to the functional median of the training data, previously normalized through GPA (or other alignment method) and allometric correction. While both methods appear robust, it is essential to understand the characteristics of the sample in relation to the variability of the individuals being examined or the potential morphological variability to be analyzed to preferentially use one method over the other. It is also imperative to understand the relationship between the recorded shape variables to explore the effect of artifacts such as collinearity, which could lead to suboptimal or unsatisfactory classification results based on the factors under study. Knowing the type of collinearity present in the data is essential for deciding the appropriate classification method to use for a given dataset.

The method presented here proposes a solution for evaluating the nutritional status of infants and children aged 6-59 months using GM techniques applied to the analysis of the left arm's shape from photographs. The overall goal of the SAM Photo Diagnosis App(c) Program is to design and develop a smartphone tool that integrates all this morphogeometric information offline and allows the training sample to be updated conveniently across different nutritional screening campaigns and contexts as new data are generated. The methodology devised and presented here, considering all these factors, offers a robust solution for the evaluation of out-of-sample cases. Maximizing the preservation of the landmark configuration used to record the arm's shape has also facilitated the explainability and interpretability of the classifier's results.

**Acknowledgments:**

The authors would like to acknowledge all the participants and families who took part in this study, as well as Action Against Hunger colleagues, field teams and everyone in Dakar and Matam offices. Special thanks to Nutrition & Health team (Fraçoise Siroma, Baye Mody Thiam and Yahya Gnokane) as well as Matam's project's field team (Seynabou Lah, Dawda Diallo, Marie Noelle Gómis). The authors also acknowledge their academic partner Université Cheikh Anta Diop, Dakar, Senegal, Ministry of Health in Senegal, CNDN and regional health authorities and staff in Matam region.
Ana Arribas Gil acknowledges funding from the Spanish Agencia Estatal de Investigación grants PID2021-123592OB-I00 and TED2021-129316B-I00.

**Author contributions:**
Conceived and designed the experiments and methods: L.M. & A.A-G. Sampling design: L.M & A.A-G. Image treatment design: L.M. Methods design: L.M., A.A-G., A.P-R. Performed the field experiments: L.M. Analyzed the data: L.M., A.A-G., A.P-R. Wrote the article: L.M. & A.A-G. Tables design: L.M. & A.A-G. Figures design: L.M. & A.A-G. Contributed to discussion: L.M., A.A-G., A.P-R., A.G.

**Data availability statement:**

The data generated and analyzed during this study are not openly available due to privacy, legal and ethical reasons. They are property of Action Against Hunger Spain (https://accioncontraelhambre.org/es). A minimal dataset and the codes necessary to reproduce the analysis presented in this work can be made available from the corresponding author upon reasonable request, in accordance with the local, national and European Union regulations. The R script allowing to reproduce the methodology described and the simulation study is available on GitHub (https://github.com/aarribasuc3m/GM_out_of_sample_classification).

# Classification of nutritional status of children from Geometric Morphometrics: an approach to out-of-sample data

Medialdea L., Arribas-Gil A., Pérez-Romero A., Gómez A.

## Supplementary information

This document contains:

A. Descriptive information about the sample analyzed for this study summarized in three tables:

1. the distribution of individuals in the sample by sex, age, and nutritional status;
2. the descriptive analysis of various collected anthropometric variables;
3. the analysis of the influence of sex, grouped age, and nutritional status on shape and size components, both before and after performing allometric correction.

B. Invariant shape analysis

    B1. Ratios of lengths

    B2. Euclidean distance matrix analysis (EDMA)

C. Preliminary comparison of alignment methods in terms of in-sample classification performance

D. A simulation study exploring the performance of the methods described in the article on synthetic data.

# A. Descriptive information about the sample

Table S1. **Sample composition**. Infants and children aged 6-59 months with an optimal nutritional condition (ONC) or severe acute malnutrition (SAM) analyzed in this work.

| Age (months) | | ONC | | | SAM | | | Total |
|---|---|---|---|---|---|---|---|---|
| | | F | M | Total | F | M | Total | |
| ≤ 24 | 6-12 | 20 | 20 | 40 | 21 | 22 | 43 | 83 |
| | 13-24 | 32 | 30 | 62 | 29 | 28 | 57 | 119 |
| | Total | 52 | 50 | 102 | 50 | 50 | 100 | 202 |
| > 24 | 25-36 | 16 | 13 | 29 | 12 | 10 | 22 | 51 |
| | 37-48 | 21 | 17 | 38 | 17 | 17 | 34 | 72 |
| | 49-59 | 16 | 23 | 39 | 22 | 24 | 46 | 85 |
| | Total | 53 | 53 | 106 | 51 | 51 | 102 | 208 |
| Total | | 105 | 103 | 208 | 101 | 101 | 202 | 410 |

Table S2. **Anthropometrical measurements descriptive analysis**. Descriptive analysis of anthropometrical measurements recorded in infants and children aged 6-59 months with an optimal nutritional condition (ONC) or severe acute malnutrition (SAM).

| | ONC | | | | | |
|---|---|---|---|---|---|---|
| | ≤ 24 | | | > 24 | | |
| Variable | Girls | Boys | Total | Girls | Boys | Total |
| Age (months) | 14.2 ± 4.8 | 14.5 ± 5.5 | 14.3^ ± 5.1 | 40.4 ± 9.3 | 42.4 ± 9.4 | 41.4^ ± 9.4 |
| Weight (kg) | 9 ± 1.2 | 9.1 ± 1.2 | 9*^ ± 1.2 | **13.4 ± 1.9** | **14.9 ± 2.2** | 14.2*^ ± 2.2 |
| Height (cm) | 75.2 ± 5.8 | 74.7 ± 6.2 | 74.9*^ ± 6 | **95.3 ± 7.4** | **99.3 ± 7.6** | 97.3*^ ± 7.8 |
| MUAC (cm) | 14.5 ± 0.8 | 14.6 ± 0.6 | 14.6*^ ± 0.7 | **15.6 ± 0.8** | **16.1 ± 1.1** | 15.8*^ ± 1 |
| WHZ | -0.4 ± 0.8 | -0.2 ± 1 | -0.3* ± 0.9 | -0.6 ± 0.5 | -0.3 ± 0.8 | -0.4* ± 0.6 |
| WAZ | -0.5 ± 0.8 | -0.6 ± 0.8 | -0.6* ± 0.8 | **-0.9 ± 0.6** | **-0.5 ± 0.9** | -0.7* ± 0.8 |
| HAZ | -0.2 ± 1.3 | -0.3 ± 1.3 | -0.2* ± 1.3 | **-0.8 ± 1.2** | **-0.1 ± 1.3** | -0.5* ± 1.3 |
| | SAM | | | | | |
| | ≤ 24 | | | > 24 | | |
| Variable | Girls | Boys | Total | Girls | Boys | Total |
| Age (months) | 13.8 ± 4.3 | 14.2 ± 5.8 | 14^ ± 5.1 | 40.6 ± 10.9 | 42.3 ± 10.4 | 41.4^ ± 10.6 |
| Weight (kg) | 6.0 ± 0.9 | 6.3 ± 1.2 | 6.2*^ ± 1.1 | 10.5 ± 1.5 | 10.7 ± 1.4 | 10.6*^ ± 1.4 |
| Height (cm) | 69.7 ± 5.3 | 69.8 ± 6.1 | 69.8*^ ± 5.7 | 95.4 ± 7.7 | 94.8 ± 7.3 | 95.1*^ ± 7.5 |
| MUAC (cm) | 11.1 ± 0.8 | 11.3 ± 1 | 11.2*^ ± 0.9 | 12.4 ± 0.8 | 12.6 ± 0.5 | 12.5*^ ± 0.7 |
| WHZ | -3.5 ± 0.8 | -3.6 ± 0.5 | -3.6* ± 0.6 | -3.4 ± 0.5 | -3.4 ± 0.4 | -3.4* ± 0.4 |
| WAZ | -3.6 ± 0.6 | -3.6 ± 0.7 | -3.6* ± 0.7 | -3.6 ± 0.6 | -3.4 ± 0.5 | -3.5* ± 0.5 |
| HAZ | **-2.0 ± 1.4** | **-2.8 ± 1.6** | -2.4*^ ± 1.5 | -1.3 ± 1.2 | -1.5 ± 1.1 | -1.4*^ ± 1.2 |

Significance level: (*) p < 0.05 for comparison between ONC and SAM children in each age group; (^) p < 0.05 for comparison between children under and over 24 months in each nutritional status group; bold for comparison between sexes in the same age and nutritional status group (p <0.05).

Table S3. **Age, sexual and nutritional effects on children's body shape and size**. Procrustes ANOVA test for shape (Procrustes coordinates) and size (log centroid size) in the whole sample by means of age groups, sex, and nutritional status. Procrustes sums of squares (SS). Procrustes mean squares (MS). degrees of freedom (df). Goodall's F statistic (F).

| | | Shape | | | | |
|---|---|---|---|---|---|---|
| Procrustes ANOVA | Effect | SS | MS | df | F | p-value |
| Before Size correction | Nutritional Status | 1,73E-01 | 4,81E-03 | 36 | 46.52** | <0.0001 |
| | Age group | 3,60E-01 | 9,99E-03 | 36 | 110.03** | <0.0001 |
| | Sex | 1,09E-03 | 3,03E-05 | 36 | 0.26 | 1 |
| After Size correction | Nutritional Status | 2,42E-01 | 6,73E-03 | 36 | 96.79** | <0.0001 |
| | Age group | 1,20E-02 | 3,34E-04 | 36 | 3.92** | <0.0001 |
| | Sex | 2,96E-03 | 8,23E-05 | 36 | 0.96 | 0.5388 |
| | | Size | | | | |
| Procrustes ANOVA | Effect | SS | MS | df | F | p-value |
| Before Size correction | Nutritional Status | 1,90E+04 | 1,90E+04 | 1 | 4.81* | 0.0288 |
| | Age group | 1,07E+06 | 1,07E+06 | 1 | 782.06** | <0.0001 |
| | Sex | 5,14E+03 | 5,14E+03 | 1 | 1.29 | 0.2565 |
| After Size correction | Nutritional Status | 0,00E+00 | 0,00E+00 | 1 | 0.12 | 0.7243 |
| | Age group | 5,10E-05 | 5,10E-05 | 1 | 33.84** | <0.0001 |
| | Sex | 1,00E-06 | 1,00E-06 | 1 | 0.52 | 0.4703 |

Significance level: (*) p < 0.05, (**) p<0.0001

## B. Invariant shape analysis

B.1 Ratios of lengths

The distribution of the ratio variables defined in Section 2.2.3 across nutrional status, sex, age and sample is presented in Figure S1.

The results of leave-one-out classification using all four ratio variables are presented in Table S4. Notice that since there is no required transformation of the ratio variables, in this case out-of-sample and in-sample classification are equivalent.

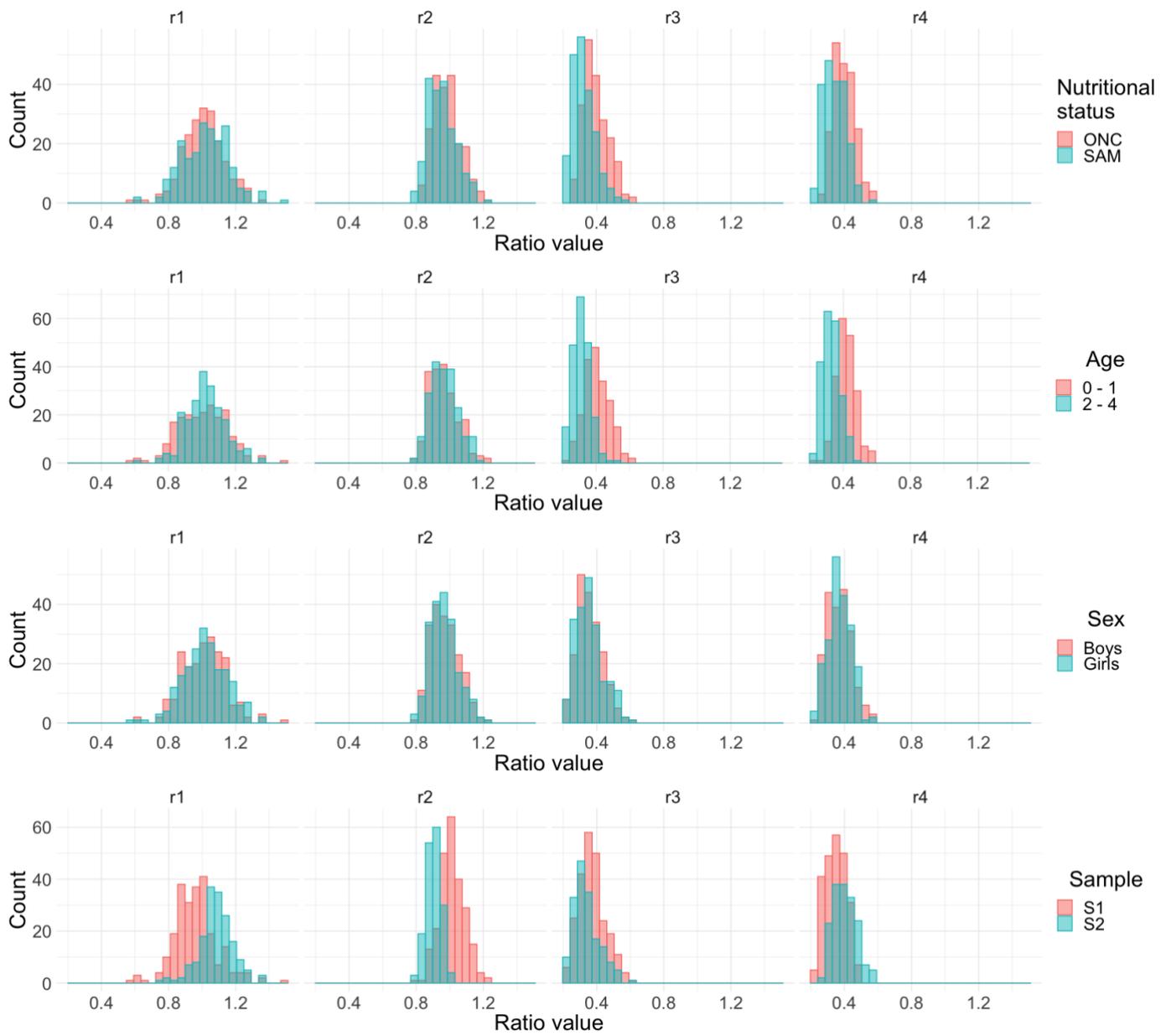

Figure S1: Histograms of ratio variables grouped by nutrional status, sex, age and subsample.

Table S4. **Classification results based on ratio variables**. Accuracy (Acc), sensibility (for SAM) and specificity (for ONC) of leave-one-out classification results using ratio variables, for Linear discriminant analysis (LDA), k-nearest neighbours (kNN, with k=5) and logistic regression (LR), in the whole sample and by age groups. The highest accuracy value for each group is highlighted in bold.

| Classification strategy | Age group | | | | | | | | |
| --- | --- | --- | --- | --- | --- | --- | --- | --- | --- |
| | Whole sample (n=410) | | | < 24 months (n=202) | | | ≥ 24 months (n=208) | | |
| | Acc | Sens | Spec | Acc | Sens | Spec | Acc | Sens | Spec |
| LDA | 0.6976 | 0.7067 | 0.6881 | **0.7970** | 0.8039 | 0.7900 | 0.8413 | 0.8302 | 0.8529 |
| LR | 0.6976 | 0.7115 | 0.6832 | 0.7871 | 0.7941 | 0.7800 | **0.8462** | 0.8491 | 0.8431 |
| kNN | **0.7000** | 0.7548 | 0.6436 | 0.7624 | 0.7843 | 0.7400 | 0.8221 | 0.8208 | 0.8235 |

B2. Euclidean distance matrix analysis (EDMA)

If we consider the two groups defined by the nutritional status in our sample, we have 202 individuals with SAM and 208 with ONC. The global shape difference test has been conducted with both the mean form matrix of the SAM group in the numerator and in the denominator. In both cases, the p-value obtained is $< 2.2 \times 10^{-16}$, indicating a significant global difference between the two groups shapes.

When conducting the local test on each of the pairs of landmarks, 81% of them showed significant differences (see Figure S2). The pairs of landmarks with the lowest ratio values are, in increasing order: 17-2, 14-13, 16-15, 12-11, 8-7, 10-9, 6-5, 18-17. All of them correspond to vertical distances between pairs of symmetrical landmarks across the longitudinal axis of the arm, that is, they account for the width of the arm (upper arm for the first ones, mainly, then forearm for the next ones, see Figure 1).

Despite of the evident differences in shape detected between the two groups, both locally and globally, classification rules to discriminate SAM from ONC based on EDMA provide similar results to ratio analysis (section B1). We have used the MDS projection from the distance between individuals defined as the logarithms of the T values. Table S5 shows the leave-one-out classification results based on the $2 \times k = 40 - dim$ MDS projection, which are slightly better than those provided with the 2-dim MDS projection (see Figure S3). In this case, the MDS projection for all individuals is obtained globally before training the classifiers, in an in-sample approach. Out-of-sample classification in this context would require the projection of each new individual into the MDS configuration defined by the remaining *n-1*.

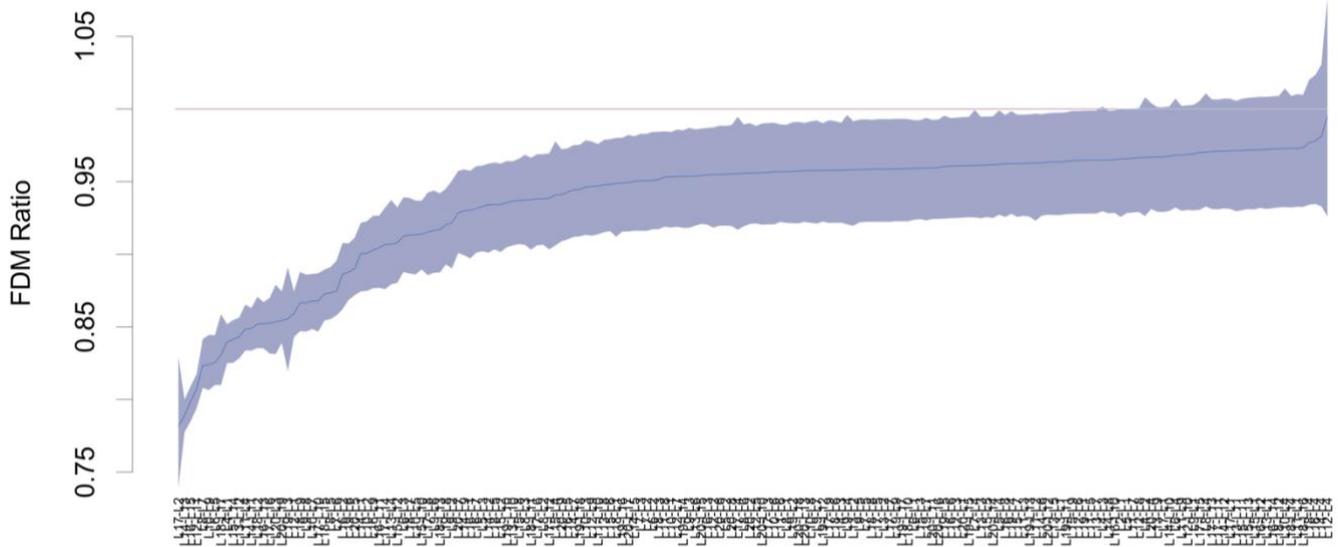

Figure S2: Bootstrap confidence intervals for the elements of the FDM obtained as the ratio between the mean form matrix of the SAM group (numerator) and the mean form matrix of the ONC group. The pairs of landmarks are arranged in the horizontal axis in increasing FDM ratio order.

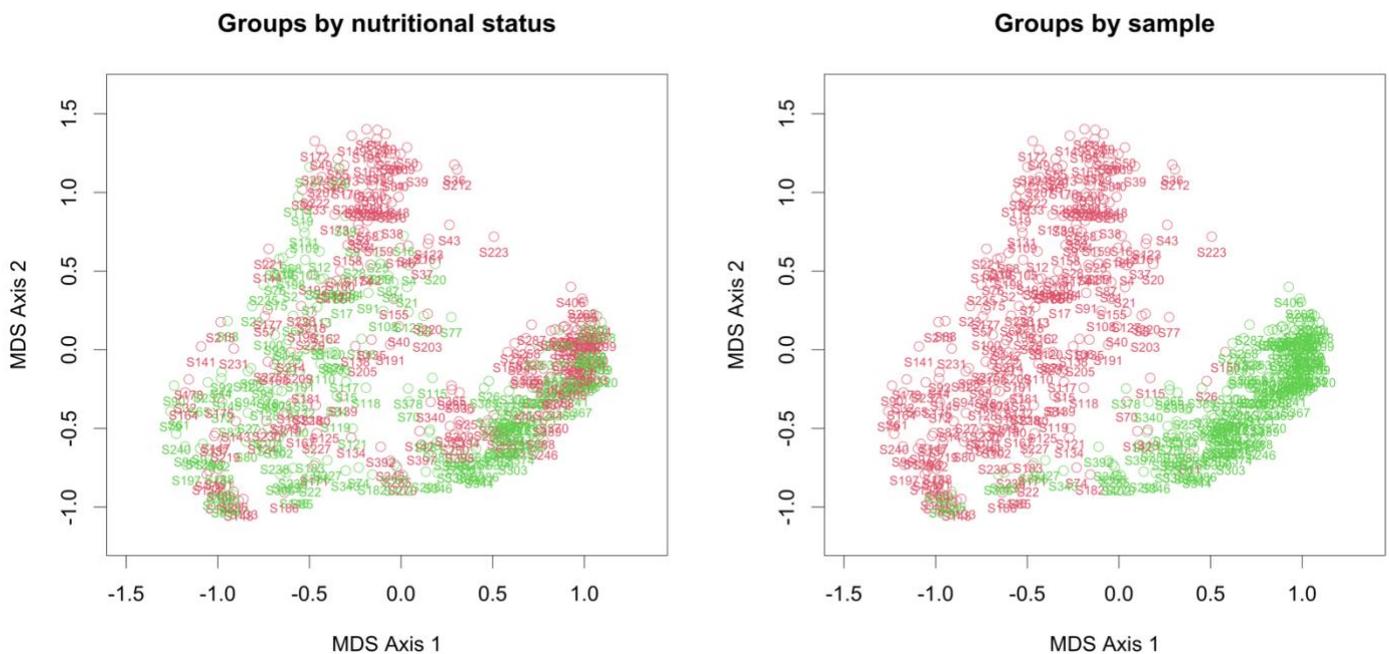

Figure S3: 2-dimensional MDS projection of the 410 individuals in the sample based on the EDMA matrix of individual distances. Left: Red bullets represent children with SAM and green bullets represent children with ONC. Right: Red bullets represent children in the first sample and green bullets represent children in the second sample.

Table S5. **Classification results based on EDMA MDS projection**. Accuracy (Acc), sensibility (for SAM) and specificity (for ONC) of leave-one-out classification results using EDMA $2 \times k = 40 - dim$ MDS projection, for Linear discriminant analysis (LDA), k-nearest neighbours (kNN, with k=5) and logistic regression (LR), in the whole sample and by age groups. The highest accuracy value for each group is highlighted in bold.

| Classification strategy | Age group ||||||||| 
|---|---|---|---|---|---|---|---|---|---|
| | Whole sample (n=410) ||| < 24 months (n=202) ||| ≥ 24 months (n=208) |||
| | Acc | Sens | Spec | Acc | Sens | Spec | Acc | Sens | Spec |
| LDA | **0.7463** | 0.7644 | 0.7277 | **0.7723** | 0.7647 | 0.7800 | **0.8317** | 0.8396 | 0.8235 |
| LR | **0.7463** | 0.7548 | 0.7376 | **0.7772** | 0.7745 | 0.7800 | 0.8125 | 0.8019 | 0.8235 |
| kNN | 0.6780 | 0.7115 | 0.6436 | 0.7030 | 0.7059 | 0.7000 | 0.7981 | 0.7925 | 0.8039 |

## C. Preliminary comparison of alignment methods in terms of in-sample classification performance

We compare three of the available alignment methods: General Procrustes Alignment (GPA), weighted Procrustes Alignment (wGPA), and Square Root Velocity Function-based alignment (SRVF). For each method, a global alignment of the entire sample is established (see Figure S4), followed by linear discriminant leave-one-out classification of each individual, training the classifier on the remaining *n-1* individuals. It is important to note that the aligned coordinates of the *n* individuals remain fixed throughout the process, as they are derived from the global alignment of the entire sample. This differs from out-of-sample classification, in which a new global alignment is obtained in each subsample of *n-1* individuals. The results are presented in Table S5, with the best classification results achieved using SRVF, which appears to be robust against allometric regression (size correction).

Table S5. **Preliminary comparison between alignment methods**. Accuracy (Acc), sensibility (for SAM) and specificity (for ONC) of leave-one-out in-sample classification results, for Linear discriminant analysis on different alignment configurations on the whole sample (n=410) obtained by Generalized Procrustes Analysis (GPA), weighted Generalized Procrustes Analysis (wGPA) and Square Root Velocity Functions (SRVF). The highest accuracy value for each classification strategy is highlighted in bold.

| Classification strategy (with LDA) | Alignment Method | | | | | | | | |
|---|---|---|---|---|---|---|---|---|---|
| | GPA | | | wGPA | | | SRVF | | |
| | Acc | Sens | Spec | Acc | Sens | Spec | Acc | Sens | Spec |
| Before size correction | 0.8000 | 0.8069 | 0.7933 | 0.7829 | 0.7871 | 0.7788 | **0.9317** | 0.9455 | 0.9183 |
| After size correction | 0.9220 | 0.9158 | 0.9279 | 0.9268 | 0.9158 | 0.9375 | **0.9317** | 0.9307 | 0.9327 |

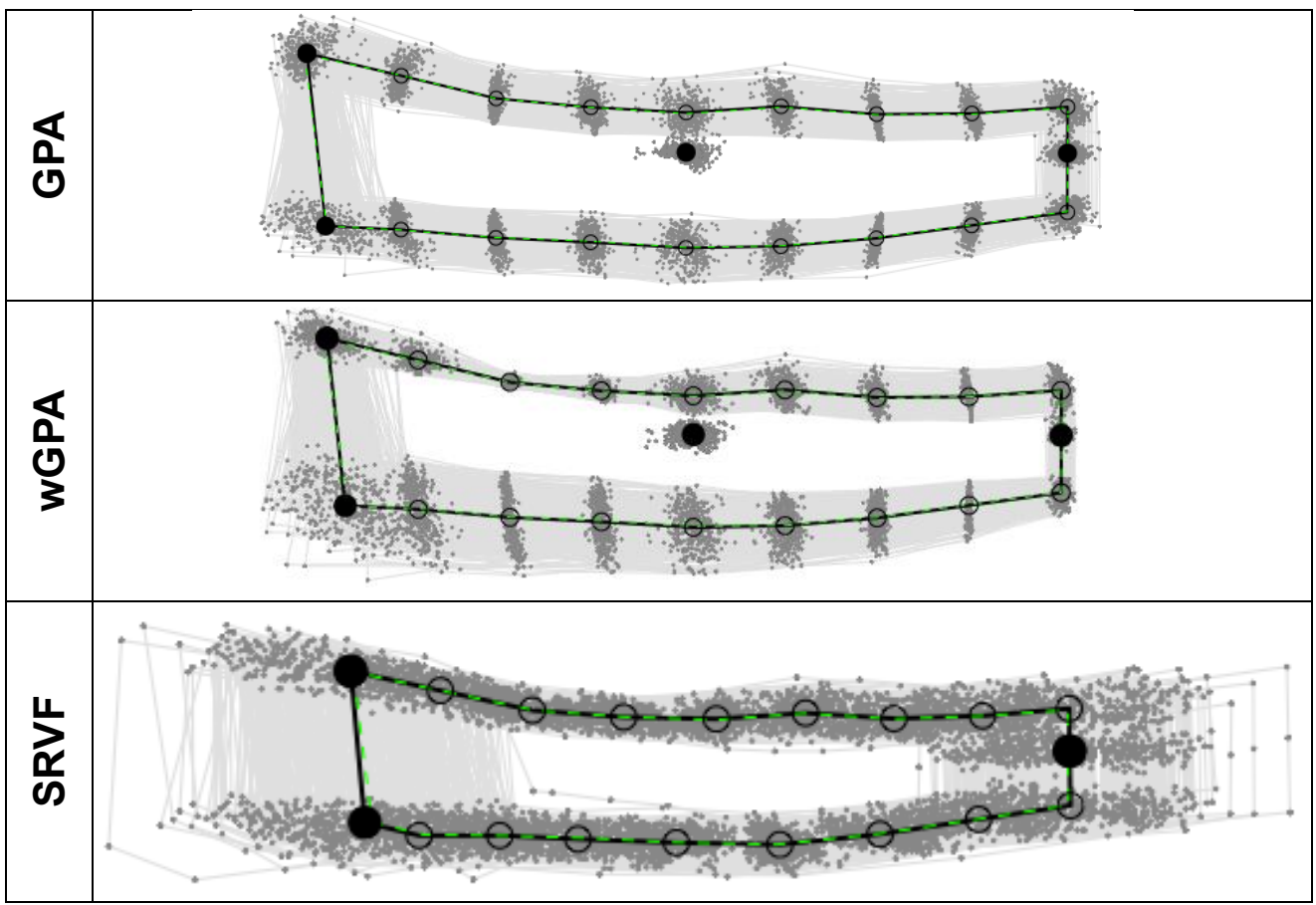

Figure S4: Aligned coordinates of the 410 individuals with three different alignment methods: Generalized Procrustes Analysis (GPA), weighted Generalized Procrustes Analysis (wGPA) and Square Root Velocity Functions (SRVF).

## D. Simulation study

In order to assess the performance of the proposed out-of-sample classification method on scenarios different than the one covered by our application, we have simulated data under five different settings. In all of them, two groups of $n = 50$ 2-dim landmarks configurations with $k = 8$ landmarks are simulated according to the specifications given below. With the aim of generating realistic data, the mean shape of two data sets of female and male gorilla skulls (`gorf.dat` and `gorm.dat` available in the R package 'shapes') have been used as mean patterns for simulation. In the following $X_1, \ldots, X_n$ denote the $n$ configuration matrices of the first group and $Y_1, \ldots, Y_n$ denote those of the second group. Each of them has dimension $k \times 2$ and they are generated as:

$$X_i = \beta_i^1 X_i^* \Gamma_i^1 + 1_k \gamma_i^{1,T} \quad \text{with} \quad X_i^* \sim MN(\mu_1, \sigma_1^2 \, I_{2 \times k}) \quad \text{i.i.d} \quad \text{for} \quad i = 1, \ldots, n$$

$$Y_i = \beta_i^2 Y_i^* \Gamma_i^2 + 1_k \gamma_i^{2,T} \quad \text{with} \quad Y_i^* \sim MN(\mu_2, \sigma_2^2 \, I_{2 \times k}) \quad \text{i.i.d} \quad \text{for} \quad i = 1, \ldots, n$$

where the values of $\mu_1, \mu_2, \sigma_1, \sigma_2$ are given below for each simulation scenario and $\beta_i^j \sim U(0,10), j = 1,2$ i.i.d, $\gamma_{i1}^j, \gamma_{i2}^j \sim U(-5,5), j = 1,2$, i.i.d, $\Gamma_i^j = \begin{pmatrix} \cos \theta_i^j & \sin \theta_i^j \\ -\sin \theta_i^j & \cos \theta_i^j \end{pmatrix}$, with $\theta_i^j \sim U(0, 2\pi), j = 1,2$ i.id.

That is, we first generate aligned configurations $X_1^*, \ldots, X_n^*$ and $Y_1^*, \ldots, Y_n^*$ and then we rotate, scale and translate each of them independently to obtain the misaligned configurations $X_1, \ldots, X_n$ and $Y_1, \ldots, Y_n$ that will be used for the analysis.

- Scenario 1: the two groups have different means but the same variation.

$$\mu_1 = ms(FG) \quad \mu_2 = ms(MG) \quad \sigma_1 = \sigma_2 = 5 \frac{1}{n} \sum_{l=1}^{k} \sum_{h=1}^{2} |\mu_{1,lh} + \mu_{2,lh}|/2$$

where $ms(FG)$ and $ms(MG)$ stand, respectively, for the mean shape of the female and male gorilla skull data sets.

- Scenario 2: the two groups have different means but the same variation, with enhanced mean difference (a translation is applied to coordinate $x$ in landmark 7 in the mean shape for the second group).

$$\mu_1 = ms(FG) \quad \mu_2 = ms(MG) + 20\lambda_{71} \quad \sigma_1 = \sigma_2 = 5 \frac{1}{n} \sum_{l=1}^{k} \sum_{h=1}^{2} |\mu_{1,lh} + \mu_{2,lh}|/2$$

where $\lambda_{71}$ is the $k \times 2$ matrix with 0's in all the entries except for the element in the 7[th] row, 1[st] column, whose value is 1.

- Scenario 3: the two groups have the same mean, but different variation, with a dispersion ratio equal to 5 between the groups.

$$\mu_1 = \mu_2 = ms(FG) \quad \sigma_1 = 5 \frac{1}{n} \sum_{l=1}^{k} \sum_{h=1}^{2} |\mu_{1,lh}| \quad \sigma_2 = 5\sigma_1.$$

- Scenario 4: the two groups have the same mean, but different variation, with a dispersion ratio equal to 3 between the groups.

$$\mu_1 = \mu_2 = ms(FG) \qquad \sigma_1 = 5 \frac{1}{n} \sum_{l=1}^{k} \sum_{h=1}^{2} |\mu_{1,lh}| \qquad \sigma_2 = 3\sigma_1.$$

- Scenario 5: the two groups have the same mean, but different variation, with a dispersion ratio equal to 1.5 between the groups.

$$\mu_1 = \mu_2 = ms(FG) \qquad \sigma_1 = 5 \frac{1}{n} \sum_{l=1}^{k} \sum_{h=1}^{2} |\mu_{1,lh}| \qquad \sigma_2 = 1.5\sigma_1.$$

For each of these scenarios, we conducted 100 simulation runs in which we applied Algorithm 2 (section 2.2.6) with allometric regression to conduct out-of-sample classification of each of the individuals of the generated sample through leave-one-out cross validation. Moreover, we applied the algorithm using three different alignment methods, namely GPA, wGPA and SRVF. The average classification results over the 100 runs are presented, respectively, in tables S6, S7 and S8. Figure S5 shows the GPA aligned samples generated under the different scenarios.

The results indicate that in all the cases LDA or kNN outperform logistic regression (except maybe in cases in which the three methods perform badly, as for Scenarios 3 to 5 with wGPA or SRVF). In the case of mean differences between groups (Scenarios 1 and 2), LDA with alignment to the mean pattern obtains the best classification results, with higher accuracy values in Scenario 2 (larger mean difference). This is particularly true when the alignment method is GPA, for which the highest accuracy values are obtained. For wGPA and SRVF alignments in Scenarios 1 and 2, the differences between using the mean and the functional median are lower. On the other hand, when the differences between the groups are only in variation (Scenarios 3,4 and 5) the classification results are in general worse than where there are differences in the mean shape. It can be observed in Figure S5 how the distorsion in the second group with respect to the mean shape is too large in Scenario 3, and that both groups are too similar in Scenario 5. However, when the alignment method used in GPA, LDA with alignment to the functional median provides satisfactory results in these challenging situations (with very good performance in Scenario 4).

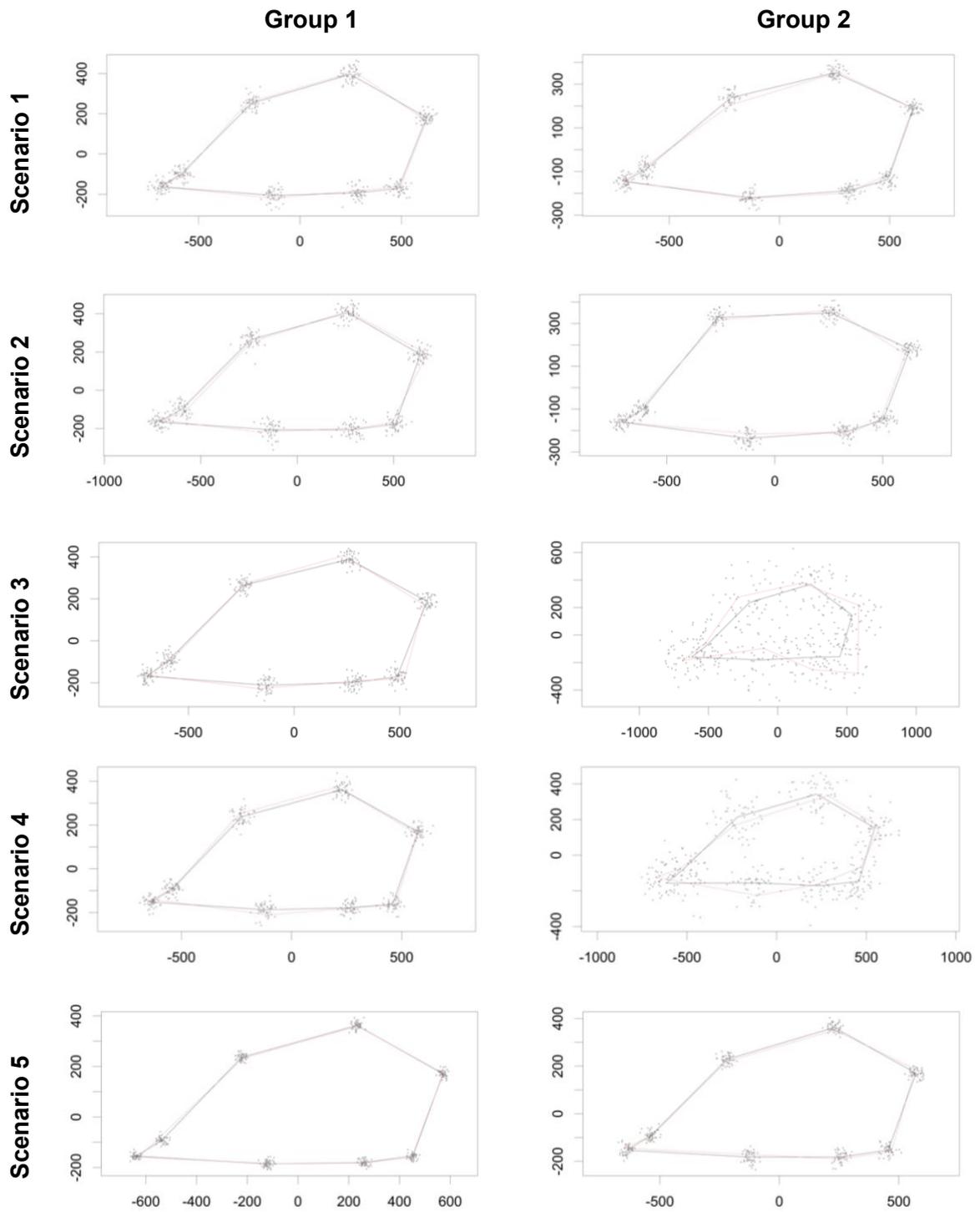

Figure S5: Data sets generated under scenarios 1 to 5. Procrustes coordinates (dark gray dots) after GPA of both groups jointly, mean (solid black) and functional median (dashed red).

Table S6. **Simulation study results, GPA alignment**. Average accuracy (Acc), sensibility (for group 1) and specificity (for group 2) of leave-one-out out-of-sample classification results over 100 simulation runs under each scenario, by Linear discriminant analysis (LDA), Logistic regression (LR) and *k*-nearest neighbours (kNN, with k=5). The highest accuracy value in each scenario is highlighted in bold.

| Class. Method | Scenario 1 | | | | | | Scenario 2 | | | | | |
|---|---|---|---|---|---|---|---|---|---|---|---|---|
| | fOPA to mean | | | fOPA to functional median | | | fOPA to mean | | | fOPA to functional median | | |
| | Acc | Sens | Spec | Acc | Sens | Spec | Acc | Sens | Spec | Acc | Sens | Spec |
| LDA | **0.8784** | 0.9208 | 0.8360 | 0.8188 | 0.8088 | 0.8288 | **0.9112** | 0.9343 | 0.8882 | 0.8152 | 0.8056 | 0.8247 |
| LR | 0.7818 | 0.9420 | 0.6216 | 0.5422 | 0.5396 | 0.5448 | 0.8167 | 0.9378 | 0.6956 | 0.5383 | 0.5370 | 0.5396 |
| kNN | 0.8591 | 0.7810 | 0.9372 | 0.8592 | 0.7804 | 0.9380 | 0.8943 | 0.8343 | 0.9542 | 0.8944 | 0.8339 | 0.9548 |

| Class. Method | Scenario 3 | | | | | | Scenario 4 | | | | | |
|---|---|---|---|---|---|---|---|---|---|---|---|---|
| | fOPA to mean | | | fOPA to functional median | | | fOPA to mean | | | fOPA to functional median | | |
| | Acc | Sens | Spec | Acc | Sens | Spec | Acc | Sens | Spec | Acc | Sens | Spec |
| LDA | 0.6334 | 0.2680 | 0.9988 | **0.8691** | 0.9268 | 0.8114 | 0.5511 | 0.1028 | 0.9994 | **0.9009** | 0.9726 | 0.8292 |
| LR | 0.5110 | 0.0228 | 0.9992 | 0.5432 | 0.5526 | 0.5338 | 0.5135 | 0.0284 | 0.9986 | 0.5097 | 0.5154 | 0.5040 |
| kNN | 0.7421 | 1 | 0.4482 | 0.6767 | 1 | 0.3534 | 0.5960 | 1 | 0.1920 | 0.5865 | 1 | 0.1730 |

| Class. Method | Scenario 5 | | | | | |
|---|---|---|---|---|---|---|
| | fOPA to mean | | | fOPA to functional median | | |
| | Acc | Sens | Spec | Acc | Sens | Spec |
| LDA | 0.5044 | 0.0098 | 0.9999 | **0.7324** | 0.6998 | 0.7650 |
| LR | 0.5115 | 0.0284 | 0.9946 | 0.5039 | 0.5032 | 0.5046 |
| kNN | 0.5562 | 0.9558 | 0.1566 | 0.5558 | 0.9570 | 0.1546 |

Table S7. **Simulation study results, wGPA alignment**. Average accuracy (Acc), sensibility (for group 1) and specificity (for group 2) of leave-one-out out-of-sample classification results over 100 simulation runs under each scenario, by Linear discriminant analysis (LDA), Logistic regression (LR) and *k*-nearest neighbours (kNN, with k=5). The highest accuracy value in each scenario is highlighted in bold.

| Class. Method | Scenario 1 | | | | | | Scenario 2 | | | | | |
|---|---|---|---|---|---|---|---|---|---|---|---|---|
| | fOPA to mean | | | fOPA to functional median | | | fOPA to mean | | | fOPA to functional median | | |
| | Acc | Sens | Spec | Acc | Sens | Spec | Acc | Sens | Spec | Acc | Sens | Spec |
| LDA | 0.8265 | 0.8064 | 0.8466 | 0.8075 | 0.7056 | 0.9094 | **0.8773** | 0.8592 | 0.8954 | 0.8617 | 0.7900 | 0.9334 |
| LR | 0.6968 | 0.7226 | 0.6710 | 0.6898 | 0.6148 | 0.7648 | 0.7372 | 0.7481 | 0.7262 | 0.7212 | 0.6798 | 0.7627 |
| kNN | **0.8454** | 0.7594 | 0.9314 | **0.8450** | 0.7604 | 0.9296 | **0.8797** | 0.8123 | 0.9472 | **0.8796** | 0.8128 | 0.9464 |

| Class. Method | Scenario 3 | | | | | | Scenario 4 | | | | | |
|---|---|---|---|---|---|---|---|---|---|---|---|---|
| | fOPA to mean | | | fOPA to functional median | | | fOPA to mean | | | fOPA to functional median | | |
| | Acc | Sens | Spec | Acc | Sens | Spec | Acc | Sens | Spec | Acc | Sens | Spec |
| LDA | 0.6214 | 0.5048 | 0.7380 | 0.6225 | 0.9988 | 0.2462 | 0.5284 | 0.5022 | 0.5546 | **0.6071** | 0.9842 | 0.2300 |
| LR | 0.4480 | 0.1268 | 0.7692 | **0.6909** | 0.9436 | 0.4382 | 0.4518 | 0.2842 | 0.6194 | **0.6025** | 0.8114 | 0.3936 |
| kNN | 0.6473 | 1 | 0.2946 | 0.6290 | 1 | 0.2580 | 0.5924 | 1 | 0.1848 | 0.5891 | 1 | 0.1782 |

| Class. Method | Scenario 5 | | | | | |
|---|---|---|---|---|---|---|
| | fOPA to mean | | | fOPA to functional median | | |
| | Acc | Sens | Spec | Acc | Sens | Spec |
| LDA | 0.5018 | 0.4932 | 0.5104 | 0.5348 | 0.7594 | 0.3102 |
| LR | 0.4982 | 0.4650 | 0.5314 | 0.5228 | 0.6514 | 0.3942 |
| kNN | 0.5620 | 0.9568 | 0.1672 | **0.5626** | 0.9580 | 0.1672 |

Table S8. **Simulation study results, SRVF alignment**. Average accuracy (Acc), sensibility (for group 1) and specificity (for group 2) of leave-one-out out-of-sample classification results over 100 simulation runs under each scenario, by Linear discriminant analysis (LDA), Logistic regression (LR) and *k*-nearest neighbours (kNN, with k=5). The highest accuracy value in each scenario is highlighted in bold.

| Class. Method | Scenario 1 | | | | | | Scenario 2 | | | | | |
|---|---|---|---|---|---|---|---|---|---|---|---|---|
| | fOPA to mean | | | fOPA to functional median | | | fOPA to mean | | | fOPA to functional median | | |
| | Acc | Sens | Spec | Acc | Sens | Spec | Acc | Sens | Spec | Acc | Sens | Spec |
| LDA | **0.7881** | 0.7620 | 0.142 | 0.7678 | 0.7518 | 0.7838 | **0.8038** | 0.7714 | 0.8363 | 0.7843 | 0.7655 | 0.8031 |
| LR | 0.7675 | 0.7770 | 0.7580 | 0.7466 | 0.7572 | 0.7360 | 0.7648 | 0.7797 | 0.7499 | 0.7451 | 0.7682 | 0.7221 |
| kNN | 0.7235 | 0.6754 | 0.7716 | 0.7196 | 0.6700 | 0.7692 | 0.7522 | 0.7181 | 0.7964 | 0.7555 | 0.7147 | 0.7963 |

| Class. Method | Scenario 3 | | | | | | Scenario 4 | | | | | |
|---|---|---|---|---|---|---|---|---|---|---|---|---|
| | fOPA to mean | | | fOPA to functional median | | | fOPA to mean | | | fOPA to functional median | | |
| | Acc | Sens | Spec | Acc | Sens | Spec | Acc | Sens | Spec | Acc | Sens | Spec |
| LDA | 0.6847 | 0.9184 | 0.4510 | 0.6860 | 0.9244 | 0.4476 | 0.6529 | 0.7864 | 0.5194 | 0.6411 | 0.7802 | 0.5020 |
| LR | **0.6915** | 0.8520 | 0.5310 | 0.6877 | 0.8592 | 0.5162 | **0.6685** | 0.7156 | 0.6214 | **0.6618** | 0.7126 | 0.6110 |
| kNN | 0.5770 | 0.9678 | 0.1862 | 0.5766 | 0.9686 | 0.1846 | 0.5307 | 0.9480 | 0.1134 | 0.5321 | 0.9488 | 0.1154 |

| Class. Method | Scenario 5 | | | | | |
|---|---|---|---|---|---|---|
| | fOPA to mean | | | fOPA to functional median | | |
| | Acc | Sens | Spec | Acc | Sens | Spec |
| LDA | **0.5433** | 0.5914 | 0.4952 | 0.5351 | 0.5646 | 0.5056 |
| LR | **0.5487** | 0.5662 | 0.5312 | **0.5419** | 0.5462 | 0.5376 |
| kNN | 0.5109 | 0.7932 | 0.2286 | 0.5139 | 0.7940 | 0.2338 |